\documentstyle[preprint,aps,eqsecnum]{revtex}

\begin{document}

%%%%%%%%%%%%%%%%%%%%%%%%%%%%%%%%%%%%%%%%%%%%%%%%%%%%%%%%%%%%%%%%
%%
%%    MACROS  (TO BE REMOVED LATER)
%%
%%%%%%%%%%%%%%%%%%%%%%%%%%%%%%%%%%%%%%%%%%%%%%%%%%%%%%%%%%%%%%%%

\def \thetad{\theta^{\dagger}}
\def \thetas{\theta^*}
\def \beq{\begin{equation}}
\def \eeq{\end{equation}}
\def \beqa{\begin{eqnarray}}
\def \eeqa{\end{eqnarray}}
\def \amp{{\cal M}}
\def \ttbar{$t \bar t$}
\def \qqbar{$q \bar q$}
\def \beamline{``beamline''}
\def \UP{{\rm R}}
\def \DOWN{{\rm L}}
\def \up{\uparrow}
\def \down{\downarrow}
\def \half{\hbox{$1\over2$}}
\def \qbar{\bar q}
\def \tbar{\bar t}
\def \ebar{\bar e}
\def \nubar{\bar\nu}
\def \trace{{\rm Tr}}
\def \nhat{{\bf{\hat{n}}}}
\def \sslash{\thinspace{\not{\negthinspace s}}}
\def \kslash{\thinspace{\not{\negthinspace k}}}
\def \pslash{\thinspace{\not{\negthinspace p}}}
\def \qslash{\thinspace{\not{\negthinspace q}}}
\def \qbarslash{\thinspace{\not{\negthinspace \qbar}}}
\def \Pslash{\thinspace{\not{\kern-1.0pt\negthinspace P}}}
\def \epsslash{\thinspace{\not{\negthinspace \epsilon}}}
\def \BD{Bj\&D}
\def \PLUS{\half(1{+}\gamma_5)}
\def \MINUS{\half(1{-}\gamma_5)}

%--------------------------------------------
%	DEFINITIONS

% Equations: abbrev., format,..
\def	\nn		{\nonumber}
\def	\=		{\;=\;}
\def	\ret		{\\[\eqskip]}
\def	\to		{\rightarrow }

% Spinor Products and Polarization Vectors
\def	\bra	{\langle}
\def 	\ket	{\rangle}
\def	\sp	#1#2{\mbox{$\bra #1 #2 \ket$}}    %spinor prod  <-+>
\def	\cp	#1#2{\mbox{$  [  #1 #2 ]$}}       %complex "    <+->
\def	\rsp	#1{\mbox{$\,\vert #1 \ket$}}      %right (ket) spinor
\def	\lsp	#1{\mbox{$\bra  #1 \!\vert\,$}}   % left (bra) spinor
\def	\sppr	#1#2{\mbox{$\bra #1 \!\vert #2 \ket$}}  %spinor prod

\draft
\preprint{
  \parbox{2in}{Fermilab--Pub--95/362-T \\
  [-0.12in] UM--TH--95--26 \\
  [-0.12in] hep-ph/9512264
}  }

\title{Angular Correlations in Top Quark Pair Production and Decay
at Hadron Colliders}
\author{Gregory Mahlon  \cite{GDMemail}}
\address{Department of Physics, University of Michigan \\
500 E. University Ave., Ann Arbor, MI  48109 }
\author{Stephen Parke \cite{SPemail}}
\address{Fermi National Accelerator Laboratory \\
P.O. Box 500, Batavia, IL  60510 }
\date{December 8, 1995}
\maketitle
\begin{abstract}
We show how to observe sizable angular correlations between the decay
products of the top quark and those of the anti-top quark in top quark
pair production and decay at hadron colliders.  These correlations
result from the large asymmetry in the rate for producing like-spin
versus unlike-spin top quark pairs provided the appropriate spin axes
are used.  The effects of new physics at production or decay on these
correlations are briefly discussed.
\end{abstract}
\pacs{}

%%%%%%%%%%%%%%%%%%%%%%%%%%%%%%%%%%%%%%%%%%%%%%%%%%%%%%%%%%%%%%%%
%%
%%      INTRODUCTION
%%
%%%%%%%%%%%%%%%%%%%%%%%%%%%%%%%%%%%%%%%%%%%%%%%%%%%%%%%%%%%%%%%%

\section{Introduction}

Now that CDF \cite{CDFtop} and D0 \cite{D0top} have observed
the top quark ($t$) and reported mass values of $176\pm8\pm10$ GeV
and $199^{+19}_{-21}\pm 22$ GeV
respectively, it is important to reconsider what other quantities
associated with top-antitop ($t\bar t\thinspace$)
production may be measured with the
data to be collected at both the Tevatron and LHC.  One interesting
avenue of investigation consists of a study of
the angular correlations
between the decay products of the top quark and those of the anti-top
quark.
For a top quark mass in the range reported by experiments,
it has been known for some time that the top quark will
decay before hadronization takes place~\cite{Bigi}.
Therefore, the angular correlations in the top quark
decay contain information on the spin of the top
quark.  If the production mechanism of the
\ttbar\ pair correlates the spins of the top and anti-top, then
this correlation will lead to angular correlations between
their decay products.

The study of angular correlations in \ttbar\ production was
pioneered by Barger, Ohnemus and Phillips~\cite{Barger1}.  These
authors concluded that the decay product angular correlations
induced by the spin correlations of top and anti-top were small
when summed over all events.
Kane {\it et. al.}~\cite{Kane2,Kane1}
reached similar conclusions in their
papers on the transverse polarization of top quarks induced
by QCD loop effects.
Since then many authors have found similar results for hadron
colliders~[\ref{hadronSTART}--\ref{hadronEND}]. Other
studies have addressed
this issue at lepton colliders~[\ref{leptonSTART}--\ref{leptonEND}].

In this paper we exploit the fact that,
even though the net polarization of top quark pairs is very small,
there is a very large asymmetry in the rate for producing
the like-spin versus unlike-spin top quark pairs
at hadron colliders if the appropriate spin axes are chosen.
Barger {\it et. al.}~\cite{Barger1} used this fact to
explain the small global correlation features of top
quark production at the Tevatron,
while Schmidt and Peskin~\cite{Schmidt} used this asymmetry
to study CP violation near threshold at the LHC and SSC.
However, this asymmetry in the number of like- to unlike-spin
top pairs is true at any hadron collider
independent of whether the top quarks are produced
via gluon-gluon fusion or quark-antiquark annihilation both
near and far from threshold.
To use the spin correlation induced by this asymmetry,
we make simple cuts on the top (anti-top) quark side
of an event to select
a given spin for the top (anti-top) quark, and then observe specific
correlations in the decay products on the anti-top (top) quark
side of the event.
These correlations are large and can be observed in the \ttbar\ events
at both the Tevatron and the LHC.

Our discussion is organized as follows.  In Sec.~\ref{PRODUCTION}
we examine the amplitudes for $q \bar q \rightarrow t \bar t$
and $gg \rightarrow t \bar t$ with polarized top quark production.
Our emphasis will be upon the excess of unlike-spin \ttbar\ pairs at
the Tevatron and the excess of like-spin \ttbar\ pairs at the LHC.
The form of the relevant amplitudes using an appropriate choice of the
spin axes and the relative parton luminosities at the two machines
combine to produce these asymmetries.
A description of the
spinor helicity basis for massive particles used in this
section appears in the Appendix, and is presented here because of its
broad applicability.
In Sec.~\ref{DECAY} we review the decay of a polarized top quark.
In Sec.~\ref{CORRELATIONS} we describe how to observe the
angular correlations arising from the production and decay
of \ttbar\ pairs.  We briefly discuss some possibilities for
new physics effects in Sec.~\ref{NEWPHYS}.
Finally, Sec.~\ref{CONCLUSIONS} contains the conclusions.

%%%%%%%%%%%%%%%%%%%%%%%%%%%%%%%%%%%%%%%%%%%%%%%%%%%%%%%%%%%%%%%%
%%
%%      POLARIZED PRODUCTION
%%
%%%%%%%%%%%%%%%%%%%%%%%%%%%%%%%%%%%%%%%%%%%%%%%%%%%%%%%%%%%%%%%%

\section{Polarized \ttbar\ Production} \label{PRODUCTION}

In this section we
present the squares of the helicity amplitudes for polarized
\ttbar\ production for both
quark-antiquark ($q \bar q\thinspace$) and gluon-gluon ($gg$)
initial states.
The expressions given below have been summed over
the spins of the initial partons, as well as
the colors of both the initial and final states.
Spin- and color-averaging factors have {\it not}\ been included.
We represent the momentum of the
particle by its symbol and decompose the top quark (anti-top quark)
momentum into a sum of two massless momenta, $t=t_1+t_2$
($\tbar =\tbar_1+\tbar_2$), such that in rest frame of the top quark
(anti-top quark) the spatial momentum of $t_1$ ($\tbar_1$) defines
the spin axis for the top (anti-top) quark (see
the appendix for details).

\newpage
For $q \bar q \rightarrow t \tbar$,
we have~\cite{Kane1}
\beqa
\sum_{\up\up{,~}\down\down}~|\amp(q\bar q \rightarrow t\bar t)|^2 =
{ {16 g^4} \over {(2q\cdot\qbar)^2} }
\Bigl[ &&
(2q\cdot t_1)(2\qbar\cdot\tbar_2)
+(2q\cdot\tbar_1)(2\qbar\cdot t_2)
\cr\noalign{\vskip8pt}
&& \enspace
+ \frac{1}{m_t^2}~\trace\thinspace
({q\thinspace t_1\thinspace t_2\thinspace
\qbar\thinspace \tbar_2\thinspace \tbar_1})
\Bigr]
   + (\thinspace q \leftrightarrow \qbar\thinspace )
\label{ANYqqlike}
%%  page t365D
\eeqa
for the production of like-spin \ttbar\ pairs, and
\beqa
\sum_{\up\down{,~}\down\up}~|\amp(q\bar q \rightarrow t\bar t)|^2 =
{ {16 g^4} \over {(2q\cdot\qbar)^2} }
\Bigl[ &&
(2q\cdot t_1)(2\qbar\cdot\tbar_1)
+ (2q\cdot\tbar_2)(2\qbar\cdot t_2)
\cr\noalign{\vskip8pt}
&& \enspace
+ \frac{1}{m_t^2}\thinspace
       \trace\thinspace
({q\thinspace t_1\thinspace t_2\thinspace
\qbar\thinspace \tbar_1\thinspace \tbar_2\thinspace})
\Bigr]
   + (\thinspace q \leftrightarrow \qbar\thinspace )
\label{ANYqqoppo}
%%  page t365E
\eeqa
for unlike-spin pairs~\cite{NOslashes}.
Note that the sum of~(\ref{ANYqqlike})
and~(\ref{ANYqqoppo}) does not depend on the
decomposition of the quark momenta.

The following expressions hold for initial
state gluons~\cite{Kane1,ggonly}:
\beqa
\sum_{\up\up{,~}\down\down} &&  ~|\amp(gg\rightarrow t\bar t)|^2 =
\cr\noalign{\vskip8pt} && \quad
{4\over3}\thinspace{g^4}
\Biggl\{ { {4}\over{(t\cdot g_1)^2} }
-{ {1}\over{(t\cdot g_1)(t\cdot g_2)} }
+{ {4}\over{(t\cdot g_2)^2} } \Biggr\}
\Biggl\{
m_t^2~[ (2t_1\cdot\tbar_1) + (2t_2\cdot\tbar_2)]
\cr\noalign{\vskip8pt}
&& \qquad\qquad
-\thinspace\thinspace {
{ \trace\thinspace( g_1 \thinspace t\thinspace  g_2\thinspace  \tbar) }
\over { (2g_1\cdot g_2)^2 }
}
\Bigl[
(2g_1\cdot t_1)(2g_2\cdot\tbar_2)
+(2g_1\cdot\tbar_1)(2g_2\cdot t_2)
\cr\noalign{\vskip8pt}
&& \qquad\qquad\qquad\qquad\qquad\quad\enspace
+\frac{1}{m_t^2}~
\trace\thinspace
({g_1\thinspace t_1\thinspace t_2\thinspace
g_2\thinspace \tbar_2\thinspace \tbar_1})
\Bigr]
\Biggr\}
+ (\thinspace g_1 \leftrightarrow g_2\thinspace ),
\label{ANYgglike}
%%  page t354B
\eeqa
and
\beqa
\sum_{\up\down{,~}\down\up} && ~|\amp(gg\rightarrow t\bar t)|^2 =
\cr\noalign{\vskip8pt} && \quad
{4\over3}\thinspace{g^4}
\Biggl\{ { {4}\over{(t\cdot g_1)^2} }
-{ {1}\over{(t\cdot g_1)(t\cdot g_2)} }
+{ {4}\over{(t\cdot g_2)^2} } \Biggr\}
\Biggl\{ m_t^2 \Bigl[
(2t_1\cdot\tbar_2) + (2\tbar_1\cdot t_2)  \Bigr]
\cr\noalign{\vskip8pt}
&& \qquad\qquad
-\thinspace\thinspace {
{ \trace\thinspace({ g_1\thinspace t\thinspace g_2\thinspace \tbar}) }
\over
{ (2g_1\cdot g_2)^2 }
}
\Bigl[
(2g_1\cdot t_1)(2g_2\cdot\tbar_1) + (2g_1\cdot\tbar_2)(2g_2\cdot t_2)
\cr\noalign{\vskip8pt}
&& \qquad\qquad\qquad\qquad\qquad\quad\enspace
+\frac{1}{m_t^2}\thinspace
\trace\thinspace
({g_1\thinspace t_1\thinspace t_2\thinspace
g_2\thinspace \tbar_1\thinspace \tbar_2})
\Bigr]
\Biggr\}
+ (\thinspace g_1 \leftrightarrow g_2\thinspace ).
\label{ANYggoppo}
%%  page t354D
\eeqa

As presented, Eqs.~(\ref{ANYqqlike})--(\ref{ANYggoppo}) are
valid for arbitrary choices of the axes along which
the $t$ and $\bar t$ spins are decomposed.
However, all choices are not equally effective for extracting
the correlations at hadron colliders.  In fact, the same choice
may not be ideal for all colliders.
We shall now describe two different bases, one of which turns
out to be well-suited to
studies at the Tevatron, while the other is
useful at both the LHC and Tevatron.

The first basis we consider is what we will refer to
as the \beamline\ basis.   It utilizes
the spin axes $p_L$ for the top quark and $p_R$ for the anti-top quark
({\it i.e.} $t_1 \propto p_L$ and $\bar t_1 \propto p_R$ in
eqs.~(\ref{ANYqqlike})--(\ref{ANYggoppo})),
where $p_L$ and $p_R$ are light-like vectors parallel to the
left and right moving beams, respectively~\cite{antibeam}.
Fortunately,
the amplitude combinations we are considering are symmetric under the
interchange of the initial parton momenta; therefore, it is not
necessary to determine the identity of each initial parton.
Furthermore, this particular basis provides a frame-independent
decomposition into like- and unlike-spin pairs.
We work in the zero momentum frame of the initial parton
pair, where we may
describe the top pair production cross section in terms of the
scattering angle $\thetas$ between the top quark and the left
moving beam, and the speed $\beta$ of the top quark.
For the \qqbar\ initial state we find
\beqa
\sum_{\up\up{,~}\down\down}~|\amp(q\bar q \rightarrow t\bar t)|^2
& = & 8g^4 \thinspace\thinspace
{
{ \beta^2(1-\beta^2)\sin^2\thetas }
\over
{ ( 1-\beta\cos\thetas )^2 }
},
\label{ZMFqqlikeB}
\\[0.10in]
\sum_{\up\down{,~}\down\up}~|\amp(q\bar q \rightarrow t\bar t)|^2
& = & 8g^4 \thinspace\thinspace
\Biggl[ \thinspace\thinspace\thinspace
1 + {
{ (1-\beta\cos\thetas-\beta^2\sin^2\thetas)^2 }
\over
{ ( 1-\beta\cos\thetas )^2 }
}
\Biggr].
\label{ZMFqqoppoB}
\eeqa
Notice the factor $\beta^2 (1-\beta^2)$ in the
like-spin pair amplitude~(\ref{ZMFqqlikeB}).
It supplies suppression of this component for both small and large
values of $\beta$.
In contrast,
the unlike-spin pair amplitude~(\ref{ZMFqqoppoB})
contains a contribution which is independent of $\beta$.

For the $gg$ initial state we define
the common spin-independent angular factor
\beq
{\cal Y}(\beta,\cos\thetas) \equiv
{
{ \thinspace 7+9\beta^2\cos^2\thetas\thinspace }
\over
{ (1-\beta^2\cos^2\thetas)^2 }
},
\label{chromosphere}
\eeq
in terms of which we have
\beqa
\sum_{\up\up{,~}\down\down}~|\amp(gg \rightarrow t\bar t)|^2
& = & {16\over3} g^4 \thinspace
{\cal Y}(\beta,\cos\thetas)
\cr&&\quad\times
(1-\beta^2)\Biggl[ \thinspace\thinspace\thinspace
1+\beta^2\cos^2\thetas + 2\beta^3\sin^2\thetas \thinspace
{
{ (\beta - \cos\thetas) }
\over
{ ( 1 - \beta\cos\thetas )^2 }
}
\Biggr],  \label{ZMFgglikeB}
%\\[0.1in]
\eeqa
\newpage
\beqa
\sum_{\up\down{,~}\down\up}~|\amp(gg \rightarrow t\bar t)|^2
& = & {16\over3} g^4 \thinspace
{\cal Y}(\beta,\cos\thetas)
\cr&&\quad\times
\beta^2\sin^2\thetas\Biggl[ \thinspace\thinspace\thinspace
1 + {
{ (1-\beta^2)^2 + ( 1 - \beta\cos\thetas - \beta^2\sin^2\thetas)^2 }
\over
{ ( 1 - \beta\cos\thetas )^2 }
}\Biggr].
\label{ZMFggoppoB}
\eeqa
Eq.~(\ref{ZMFgglikeB}) shows that the like-spin pairs
coming from gluon-gluon fusion will
be suppressed for large $\beta$, while~(\ref{ZMFggoppoB})
tells us that unlike-spin pairs are disfavored at low $\beta$.

%%%%%%%%%%%%%%%%%%%%%%%%%%%%%%%%%%%%%%%%%%%%%%%%%%%%%%%%%%%%%%%%%%%%%
%
%As an alternative to the above, we examine the choice
%$t_2 = \bar t_2 \propto p_L$, the ``anti-beamline'' basis.
%In this case the ZMF \qqbar\ amplitudes take the form
%\beqa
%\sum_{\up\up{,~}\down\down}~|\amp(q\bar q \rightarrow t\bar t)|^2
%& = & 8g^4 \thinspace\thinspace
%{
%{ 2(1-\beta^2) }
%\over
%{ 1-\beta^2\cos^2\thetas }
%}, \\[0.10in]
%\label{ZMFqqlikeC}
%\sum_{\up\down{,~}\down\up}~|\amp(q\bar q \rightarrow t\bar t)|^2
%& = & 8g^4 \thinspace\thinspace
%\beta^2\sin^2\thetas\thinspace\thinspace
%{
%{ 1+ \beta^2\cos^2\thetas }
%\over
%{ 1-\beta^2\cos^2\thetas }
%},
%\label{ZMFqqoppoC}
%\eeqa
%while the $gg$ amplitudes read
%\beqa
%\sum_{\up\up{,~}\down\down}~|\amp(gg \rightarrow t\bar t)|^2
%& = & {16\over3} g^4 \thinspace
%{\cal Y}(\beta,\cos\thetas)
%~{
%{4\beta^2(1-\beta^2)\sin^2\thetas}
%\over
%{ 1-\beta^2\cos^2\thetas }
%},   \\[0.1in]
%\label{ZMFgglikeC}
%\sum_{\up\down{,~}\down\up}~|\amp(gg \rightarrow t\bar t)|^2
%& = & {16\over3} g^4 \thinspace
%{\cal Y}(\beta,\cos\thetas)
%~\bigl[(1-\beta^2)^2 + \beta^4\sin^4\thetas \bigr] \thinspace
%{
%{ 1+\beta^2\cos^2\thetas }
%\over
%{ 1-\beta^2\cos^2\thetas }
%}.
%\label{ZMFggoppoC}
%\eeqa
%
%%%%%%%%%%%%%%%%%%%%%%%%%%%%%%%%%%%%%%%%%%%%%%%%%%%%%%%%%%%%%%%%%%%%%

The other basis we wish to discuss is built upon the helicities
of the $t$ and $\bar t$.  The helicity of a massive particle
is a frame-dependent concept, so the  decomposition
into like- and unlike-helicity pairs will depend upon which
frame is used.
We choose to measure the helicities of the top and anti-top quarks
in the zero momentum frame of the initial parton pair.
For the $q\bar q$ initial state
we find
\beqa
\sum_{LL{,~}RR}~|\amp(q\bar q \rightarrow t\bar t)|^2
& = & 8g^4 ~(1-\beta^2)\sin^2\thetas ,
\label{ZMFqqlike}
%%  page t432
\\[0.1in]
\sum_{LR{,~}RL}~|\amp(q\bar q \rightarrow t\bar t)|^2
& = & 8g^4 ~(1+\cos^2\thetas).
\label{ZMFqqoppo}
%%  page t432
\eeqa
We see from~(\ref{ZMFqqlike}) that in the high
energy limit ($\beta\rightarrow 1$), the production of like-helicity
\ttbar\ pairs is suppressed.

The expressions involving initial state gluons are only slightly
more complex:
\beqa
\sum_{LL{,~}RR}~|\amp(gg \rightarrow t\bar t)|^2
& = & {16\over3} g^4 \thinspace\thinspace
{\cal Y}(\beta,\cos\thetas)
{}~(1-\beta^2)(1+\beta^2 +\beta^2\sin^4\thetas),
\label{ZMFgglike}
%%  page t418
\\[0.1in]
\sum_{LR{,~}RL}~|\amp(gg \rightarrow t\bar t)|^2
& = & {16\over3} g^4 \thinspace\thinspace
{\cal Y}(\beta,\cos\thetas)
{}~\beta^2\sin^2\thetas(1+\cos^2\thetas).
\label{ZMFggoppo}
%%  page t419
\eeqa
Once again, we see suppression of like-helicity \ttbar\ pairs in the
high energy limit.  However, we note that for low energies,
unlike-helicity pair production is suppressed relative to the
production of like-helicity pairs by a factor of $\beta^2$.

\begin{figure}[h]

\vspace*{15cm}
\includegraphics{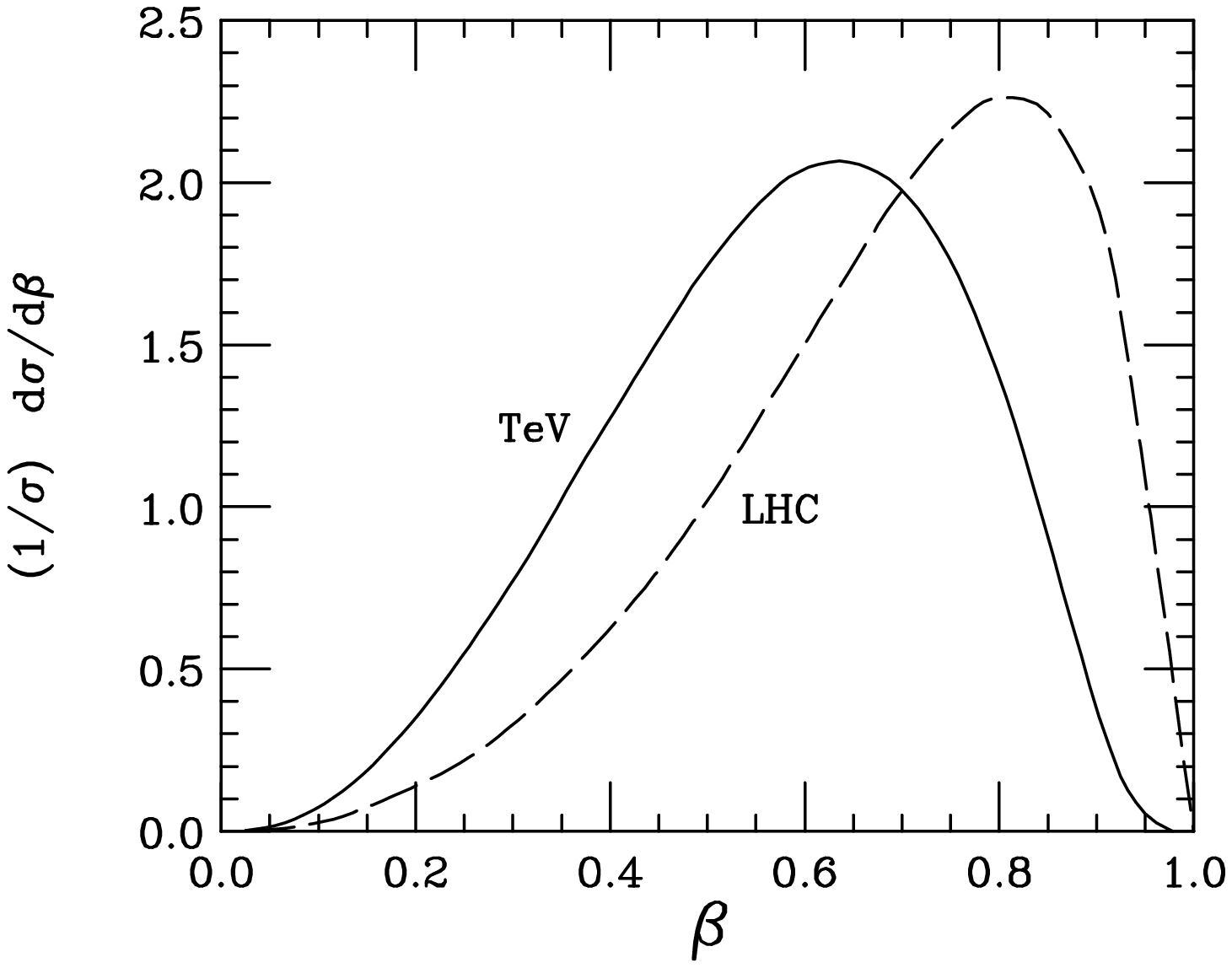}
\vspace{1.0cm}

\caption[]{Differential cross section for \ttbar\ production as
a function of the zero momentum frame speed $\beta$ of the top quark
for the 2.0 TeV Tevatron (solid) and 14 TeV LHC (dashed).}
\label{BetaPlot}
\end{figure}

\begin{figure}[h]

\vspace*{15cm}
\includegraphics{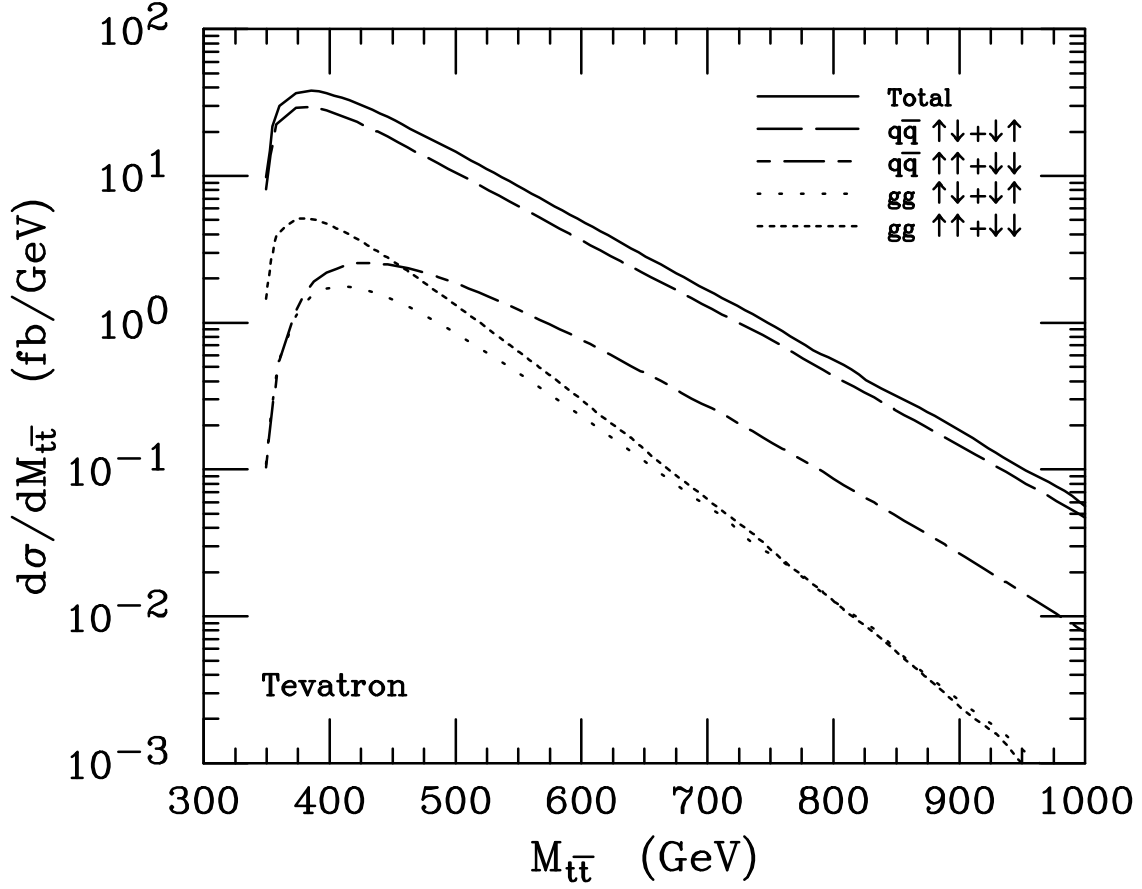}
\vspace{1.0cm}

\caption[]{Differential cross section for \ttbar\ production as
a function of the \ttbar\ invariant mass, $M_{t\bar t}$,
for the Tevatron with center of mass energy 2.0 TeV,
decomposed into
${\up\down}{+}{\down\up}$
and
${\up\up}{+}{\down\down}$
spins of the \ttbar\ pair using the \beamline\ basis
for both \qqbar\ and $gg$ components.}
\label{TeVmassplotb}
\end{figure}

\begin{figure}[h]

\vspace*{15cm}
\includegraphics{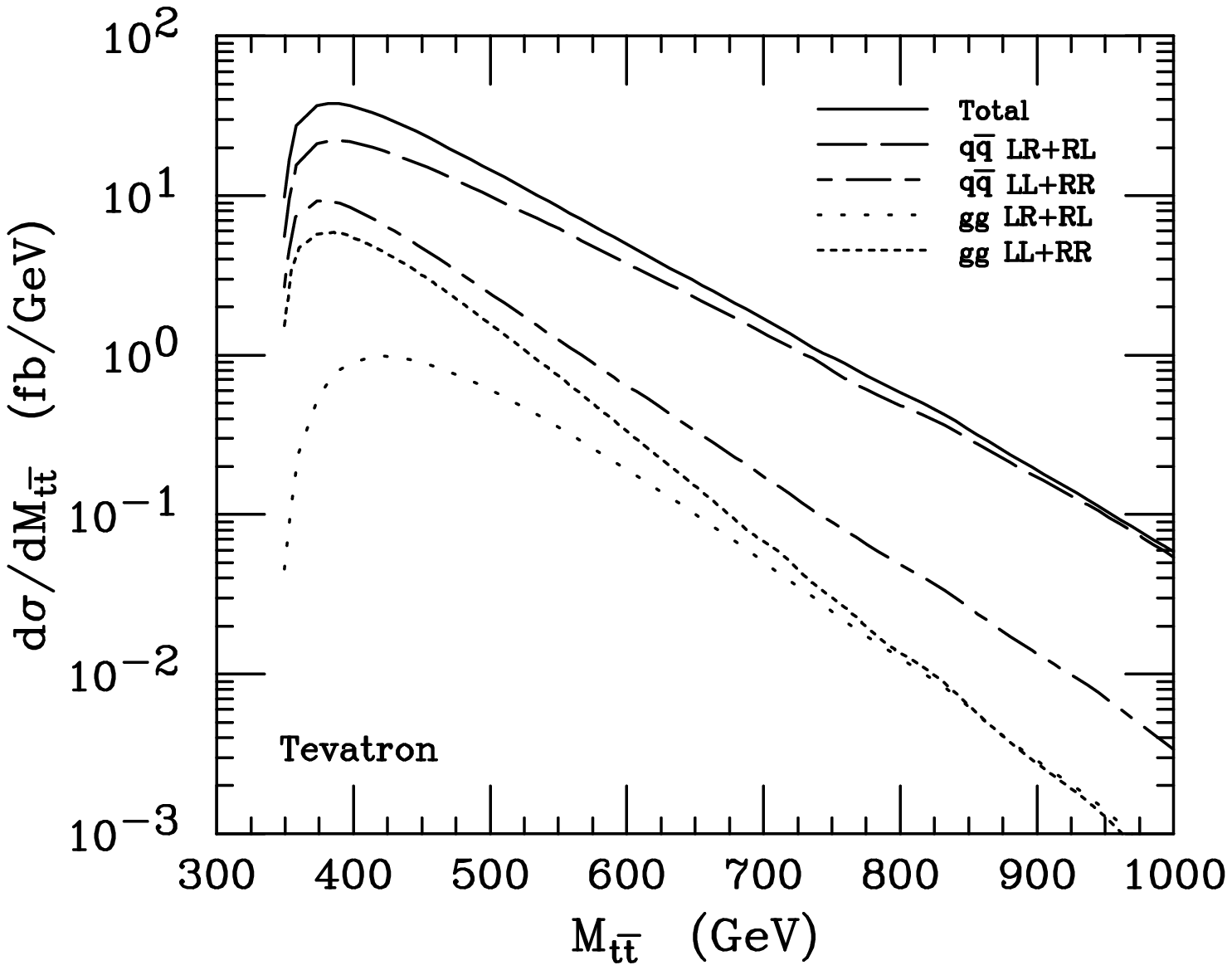}
\vspace{1.0cm}

\caption[]{Differential cross section for \ttbar\ production as
a function of the \ttbar\ invariant mass, $M_{t\bar t}$,
for the Tevatron with center of mass energy 2.0 TeV,
decomposed into LR+RL and LL+RR helicities
in the zero momentum frame of the \ttbar\ pair for both
\qqbar\ and $gg$ components.}
\label{TeVmassploth}
\end{figure}

\begin{figure}[h]

\vspace*{15cm}
\includegraphics{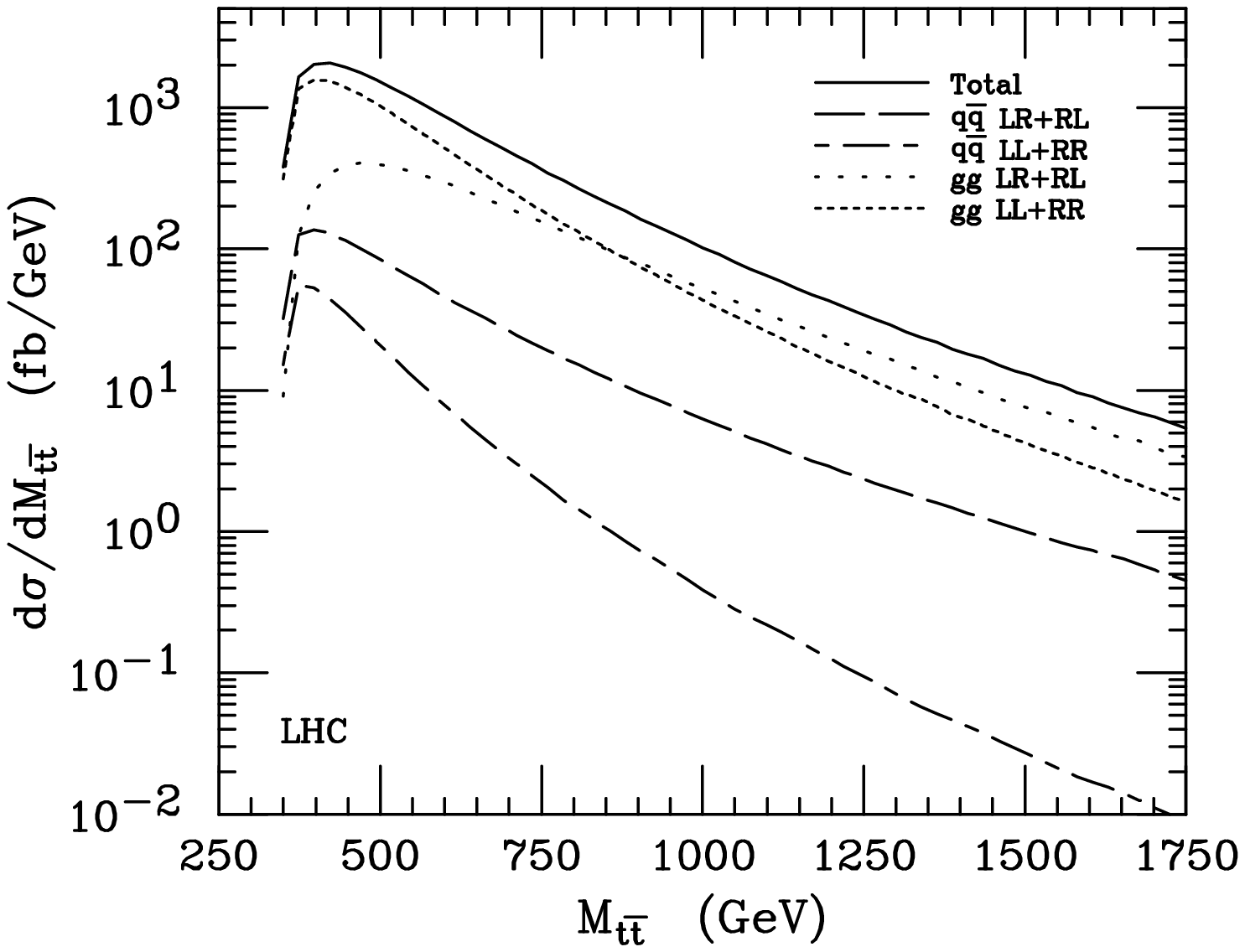}
\vspace{1.0cm}

\caption[]{Differential cross section for \ttbar\ production as
a function of the \ttbar\ invariant mass, $M_{t\bar t}$,
for the LHC with center of mass energy 14 TeV,
decomposed into LR+RL and LL+RR helicities
in the zero momentum frame of the \ttbar\ pair for both
\qqbar\ and $gg$ components.}
\label{LHC14massplot}
\end{figure}

The difference in the $\beta$ dependence of these squared matrix
elements is such that
at nearly all hadron colliders, the \ttbar\ pairs are
produced with one or other of
the spin configurations dominating the cross section.
In Fig.~\ref{BetaPlot} we show the $\beta$ distributions
for 175~GeV
top quarks produced at the Tevatron and the LHC~\cite{structfun}.
The breakdown of the total \ttbar\ cross section
into like- and unlike-spin pairs as a function of the \ttbar\
invariant mass is given in
Figs.~\ref{TeVmassplotb} and~\ref{TeVmassploth}
for the Tevatron using the \beamline\ and helicity bases respectively.
In the \beamline\ basis 80\% of the \ttbar\ pairs have unlike spins,
while in the helicity basis 67\% of the \ttbar\ pairs have
unlike helicities~\cite{stelzer}.
Fig~\ref{LHC14massplot} is the same breakdown
at the LHC using the helicity basis,
where 67\% of the \ttbar\ pairs have like helicities\cite{contrast}.
These asymmetries may be understood in terms of the
amplitudes~(\ref{ZMFqqlikeB})--(\ref{ZMFggoppo})
and relative parton luminosities at the two machines.
It is well-known that \ttbar\ production at the Tevatron is
dominated by the \qqbar\ initial state.
Furthermore, Eq.~(\ref{ZMFqqlikeB})
tells us that the production
of like-spin \ttbar\ pairs in the \beamline\ basis
from a \qqbar\ initial state is disfavored.
Consequently, most of the \ttbar\ pairs produced at the
Tevatron have unlike spins in this description.
Similar considerations may be applied to understand the production
asymmetries in terms of the helicity basis at both machines.

\begin{figure}[h]

\vspace*{15cm}
\includegraphics{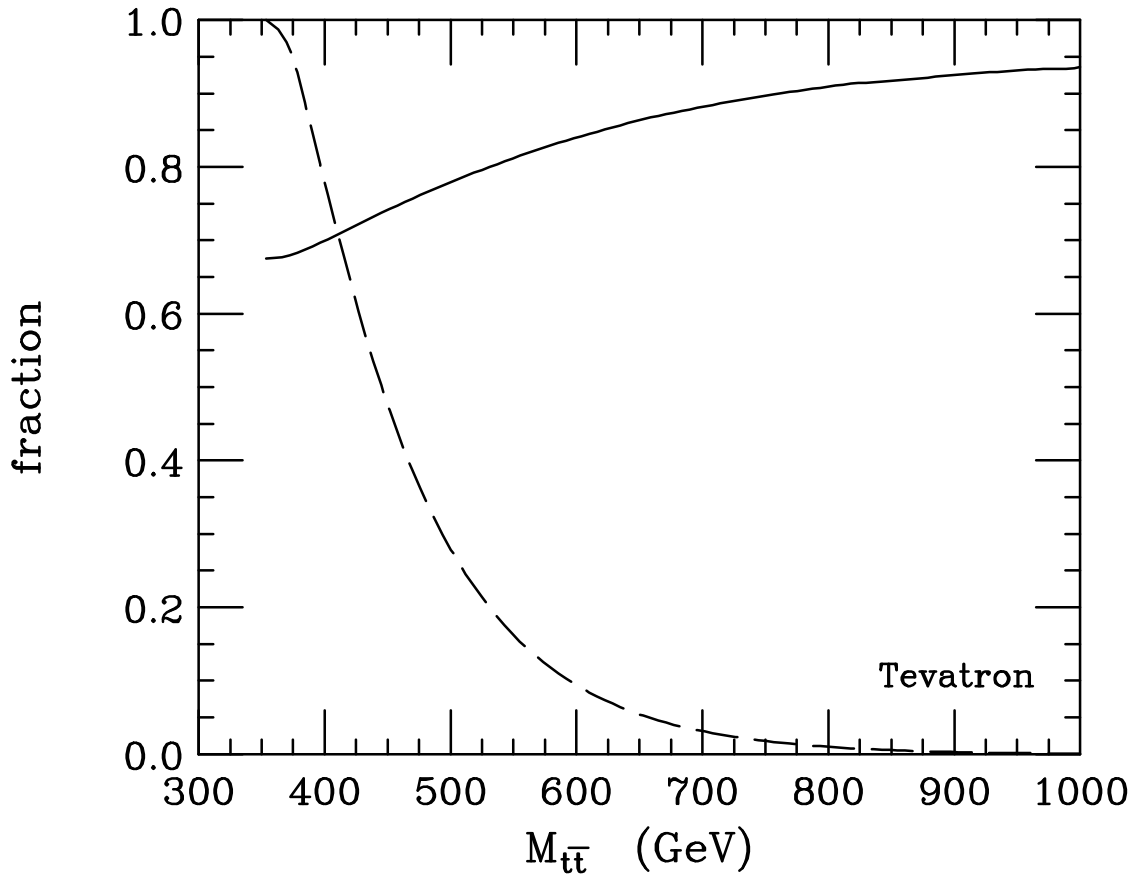}
\vspace{1.0cm}

\caption[]{
The solid curve is the fraction of those \ttbar\ pairs at the
Tevatron (2.0 TeV) with an invariant mass {\it above} $M_{t\bar t}$
which have helicities LR$+$RL.
The dashed curve is the fraction of the total cross section with an
invariant mass {\it above}\ $M_{t\bar t}$.}
\label{fracTeV}
\end{figure}

Since the $\beta$ and $\thetas$ dependence is different for
different spin configurations, we may ask if it is possible
to devise a set of cuts which would increase the purity of
the dominant spin configuration.  For the Tevatron using the
\beamline\ basis, this turns out to be difficult.  We have
found that in order to increase the fraction of unlike-spin
\ttbar\ pairs by more than a percent or two, it is necessary
to apply such stringent cuts that the statistics are reduced
by a factor of 10 or more.  Fortunately, 80\% purity is already
sufficiently good to render the correlations we wish to consider
visible (see Sec.~\ref{CORRELATIONS}).
On the other hand, using the helicity basis
at the Tevatron and requiring $M_{t\bar t}$ to be larger
than some value will improve the unlike-helicity purity
of the sample.
In Fig.~\ref{fracTeV} we show how such a cut affects
the fraction of unlike-helicity pairs, as well
as the fraction of the total \ttbar\ sample retained by such
a cut.
Using this basis with the cut $M_{t\bar{t}} > 450$ GeV
increases the unlike-helicity fraction to 74\%,
while retaining 47\% of the data sample.

\begin{figure}[h]

\vspace*{15cm}
\includegraphics{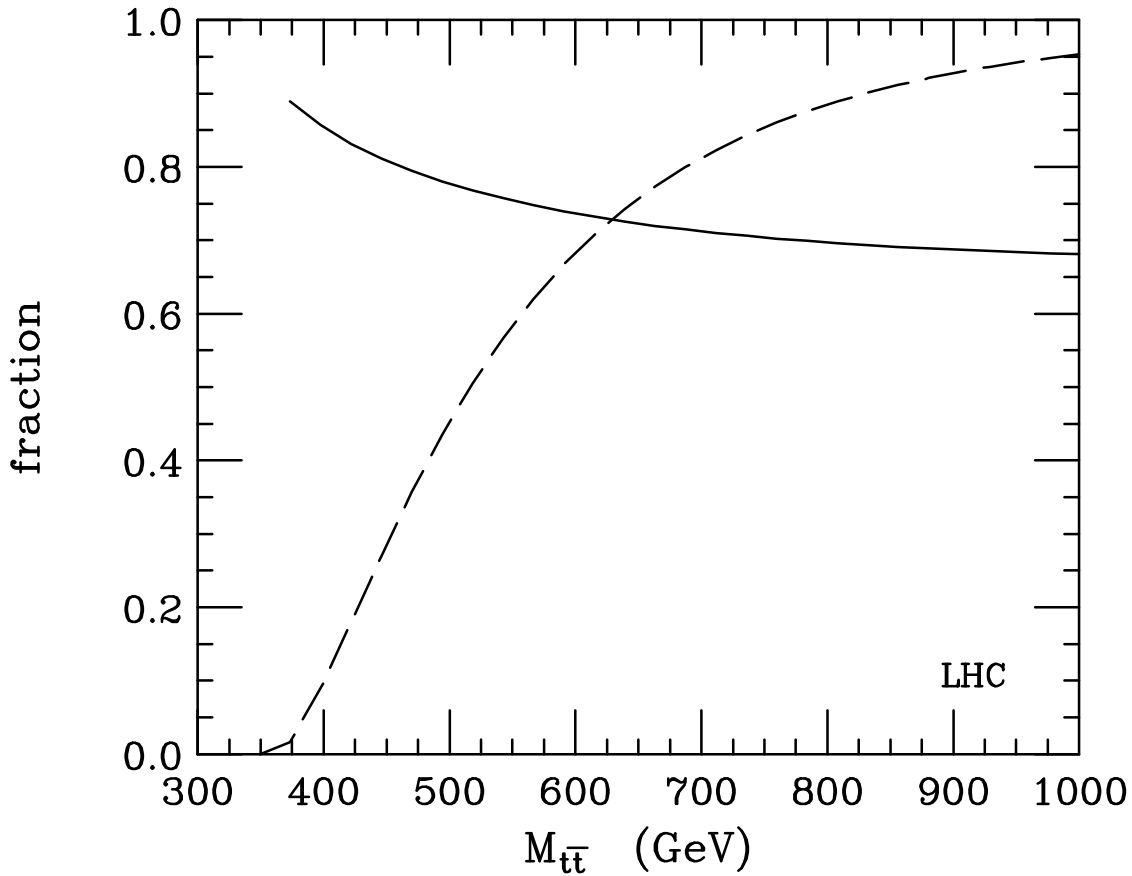}
\vspace{1.0cm}

\caption[]{
The solid curve is the fraction of those \ttbar\ pairs at the
LHC (14 TeV) with an invariant mass {\it below} $M_{t\bar t}$
which have helicities LL$+$RR.
The dashed curve is the fraction of the total cross section with an
invariant mass {\it below}\ $M_{t\bar t}$.}
\label{fracLHC}
\end{figure}

It may also be desirable at the LHC to impose a cut on $M_{t\bar t}$.
Recall that Eqs.~(\ref{ZMFgglike}) and~(\ref{ZMFggoppo})
predict that for low values of $\beta$, mostly like-helicity pairs
are produced, while for high values of $\beta$, mostly unlike-helicity
pairs are produced.  This feature is clearly visible in
Fig.~\ref{LHC14massplot}:  in the 800--900~GeV region, the
like- and unlike-helicity contributions from $gg$ become equal.
Thus, it is reasonable to consider selecting events with
$M_{t\bar t}$ less than some maximum value.
In Fig.~\ref{fracLHC} we show how such a cut affects
the fraction of like-helicity pairs, as well
as the fraction of the total \ttbar\ sample retained by such
a cut.  For example, if we impose the cut $M_{t\bar t} < 500$~GeV,
we increase the like-helicity fraction to 78\%, while retaining
45\% of the data sample.

Lastly, all of the above fractions depend only
weakly upon the value of the top quark mass, varying by only a few
percent over the range 150 GeV $< m_t <$ 200 GeV.

%%%%%%%%%%%%%%%%%%%%%%%%%%%%%%%%%%%%%%%%%%%%%%%%%%%%%%%%%%%%%%%%
%%
%%      POLARIZED DECAY
%%
%%%%%%%%%%%%%%%%%%%%%%%%%%%%%%%%%%%%%%%%%%%%%%%%%%%%%%%%%%%%%%%%

\section{Polarized Top Quark Decay} \label{DECAY}

Because of its extremely short lifetime, the top quark decays
before it hadronizes, imparting its spin
information to its decay products.
The squared matrix element for the complete decay chain is
rather simple, considering the three-body final state.
Again we decompose the top quark momentum into
two massless momenta, $t=t_1+t_2$, such that
the spatial momentum of $t_1$
defines the spin axis in the top quark rest frame.
For a top quark ($t$) decaying into a
$b$-quark ($b$), positron ($\bar e$)
and neutrino ($\nu$), we obtain
\beqa
|\amp_{\up}(t \rightarrow b\bar e\nu_e)|^2
& = &
{
{ g_w^4 (2\nu\cdot b)(2\bar e \cdot t_2) }
\over
{ (2\nu\cdot\bar e-M_W^2)^2 + M_W^2 \Gamma_W^2 }
},
\label{tRdecay}
\\[0.1in]
|\amp_{\down}(t \rightarrow b\bar e\nu_e)|^2
& = &
{
{ g_w^4 (2\nu\cdot b)(2\bar e \cdot t_1) }
\over
{ (2\nu\cdot\bar e-M_W^2)^2 + M_W^2 \Gamma_W^2 }
}.
\label{tLdecay}
\eeqa
For the hadronic decay of the top quark, $t \rightarrow b ~\bar d ~u$,
one should replace the $\bar e$ with
$\bar d$ and $\nu$ with $u$ in the above expressions.

\begin{figure}[h]

\vspace*{15cm}
\includegraphics{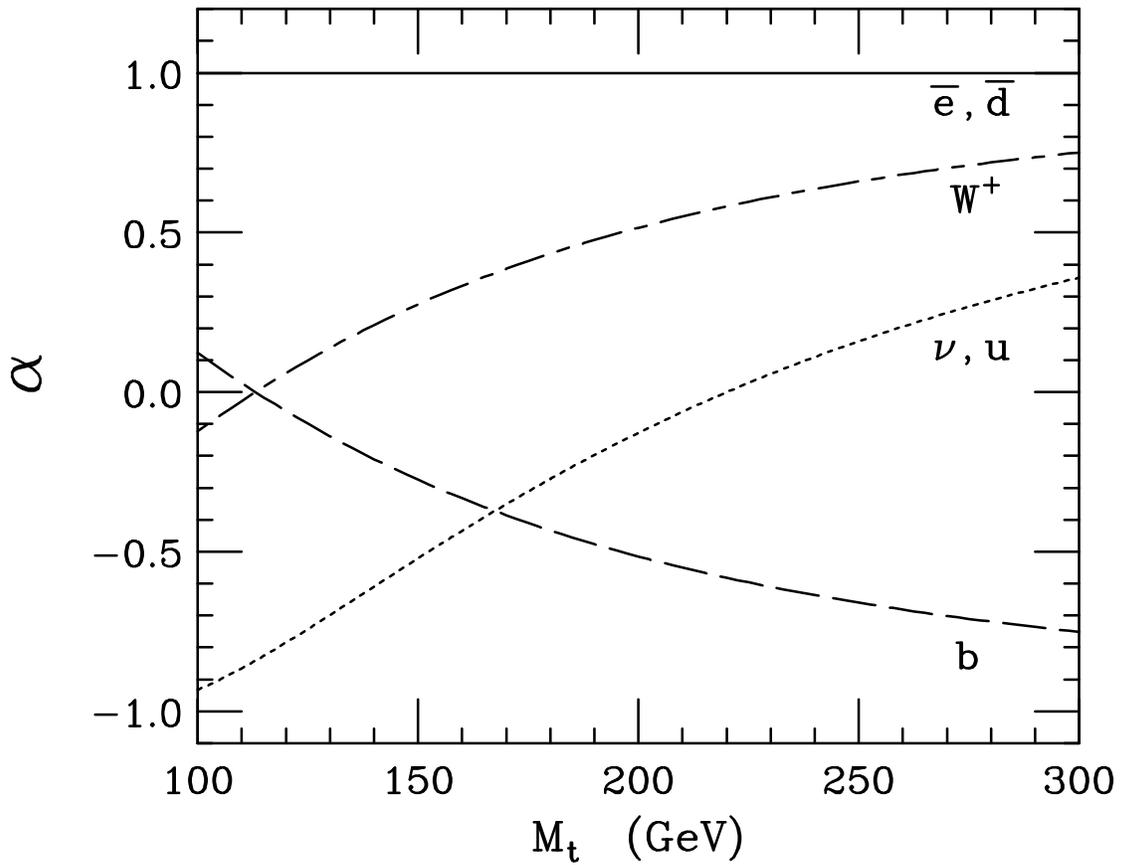}
\vspace{1.0cm}

\caption[]{Correlation coefficients, $\alpha_i$, for a spin up
top quark as a function of $m_t$, see Table~\ref{Alphas}.
\qquad\qquad\qquad\qquad\qquad}
\label{AlphaPlot}
\end{figure}

The differential decay rates in the rest frame of the
decaying particle may be parameterized as
\beq
{1 \over \Gamma}
{{d\Gamma}
\over
{d(\cos\theta_i)}
}
= { {1 ~+~ \alpha_i \cos\theta_i}\over{2} }
\label{alphadefn}
\eeq
where $\theta_i$ is the angle between the chosen spin axis
and the direction of motion of the $i$\/th decay product,
$i$ = $b$, $\bar e$, or $\nu$ (alternatively $b$, $\bar d$ or $u$).
The correlation coefficient
$\alpha_i$ may be computed from the matrix elements
(\ref{tRdecay})--(\ref{tLdecay}), see Ref. \cite{Jezabek2}.
For a spin-up top quark the results are given
in Table~{\ref{Alphas}}, and plotted in Fig.~{\ref{AlphaPlot}}.
The spin-down top quark has correlation coefficients opposite
in sign to the spin-up case,
whereas for the anti-top quark the correlation coefficients for
spin-up (spin-down) equal the coefficients for the top quark with
spin-down (spin-up).
For $m_t=175~{\rm GeV}$ the values of $\alpha_{\bar{e}},~\alpha_{\nu}$
and $\alpha_b$ for a spin-up top quark are 1, $-0.31$, and $-0.41$
respectively.

These correlations can be used to determine probabilistically
whether the top quark is spin-up or spin-down.
For the $i$\/th decay product, if $\cos \theta_i > y $ then the
probability that the top quark had spin-up, $P_{\uparrow}$,
is given by
\beq
[2~+~\alpha_i (1+y)]/4.
\label{Rcut}
\eeq

\begin{figure}[h]

\vspace*{15cm}
\includegraphics{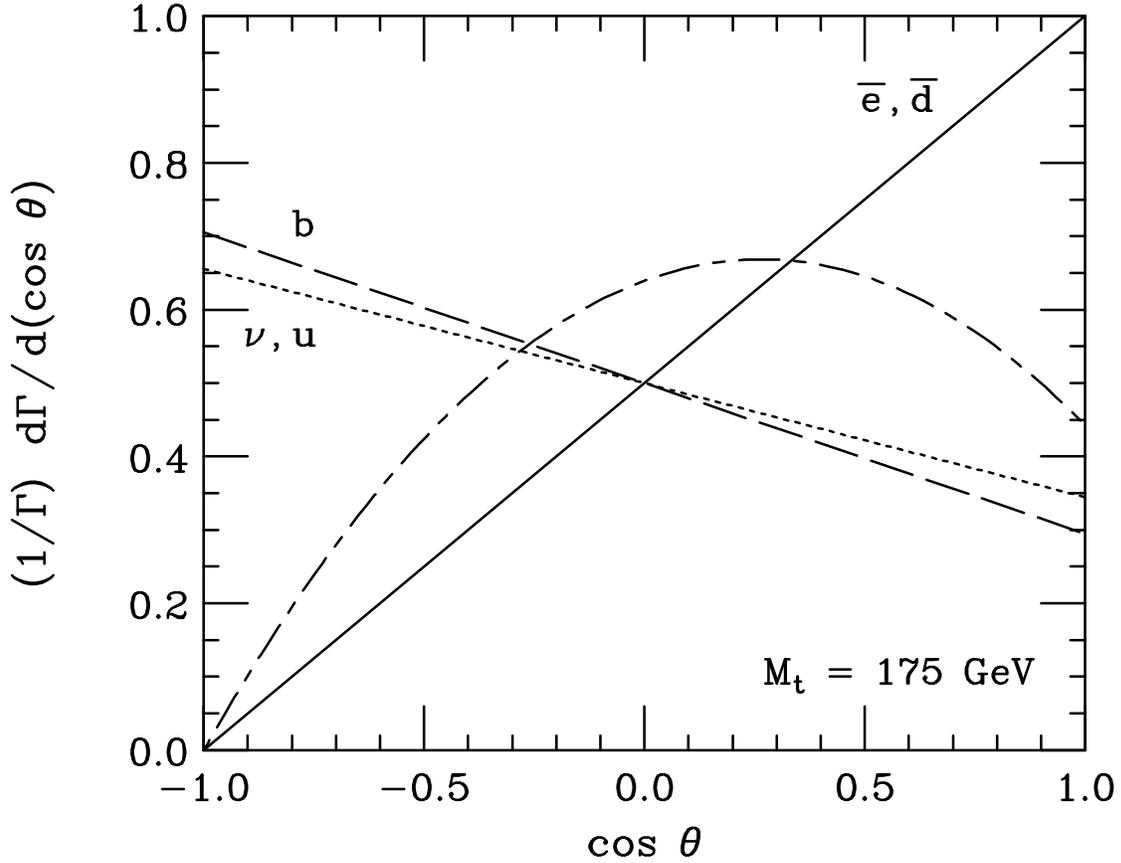}
\vspace{1.0cm}

\caption[]{Angular correlations in the decay of a 175~GeV spin-up top
quark.  The
lines labelled $\bar e$, $\bar d$, $b$, $\nu$ and $u$ are the
angle between the spin axis and the particle in the rest frame
of the top quark.  The unlabelled dot-dash line is the angle between
the $b$ quark and the $\bar e$ or $\bar d$ in the rest frame
of the $W$-boson.}
\label{tUPcorr}
\end{figure}

In the rest frame of the $W$-boson it is well-known that the
correlation of the angle, $\thetad$,
between the charged lepton (or down-type quark) and
the $b$-quark direction is given
by
\beq
{1 \over \Gamma}
{{d\Gamma}
\over
{d(\cos\thetad)}
}
= { 3\over4 } \thinspace
{ {\thinspace m_t^2 \sin^2 \thetad ~+~
 m_W^2 (1+\cos \thetad)^2}
\over
{m_t^2 + 2m_W^2} },
\label{wboson}
\eeq
reflecting the relative rate of longitudinal to transverse
$W$-bosons in top decay of $m_t^2$ to $2m_W^2$.
This correlation can be used to distinguish the $d$-type quark
from the $u$-type quark in hadronic top quark decays.
If we choose events such that one of the jets
has $\cos \thetad > z $,
then the probability that this jet orginates from
a $d$-type quark, $P_d$, is
\beq
{
{m_t^2(2-z-z^2)~+~m_W^2(7+4z+z^2)}
\over
{2~[m_t^2(2-z-z^2)~+~m_W^2(4+z+z^2)]}
}.
\label{Dcut}
\eeq
In Fig.~\ref{tUPcorr} we have plotted all the angular correlations
for a spin up top quark decay.

%%%%%%%%%%%%%%%%%%%%%%%%%%%%%%%%%%%%%%%%%%%%%%%%%%%%%%%%%%%%%%%%
%%
%%      CORRELATION
%%
%%%%%%%%%%%%%%%%%%%%%%%%%%%%%%%%%%%%%%%%%%%%%%%%%%%%%%%%%%%%%%%%

\section{Correlations in \ttbar\ Production and Decay}
\label{CORRELATIONS}

In this section we put together the spin correlations induced by
production, Sec.~{\ref{PRODUCTION}}, and the polarized decays,
Sec.~{\ref{DECAY}}. For the {\it i\/}th decay product of the top
quark with angle $\theta_i$ to the spin axis of
the top quark in the top rest frame
and the {$\bar{\imath}\kern1.0pt$}th decay product of the anti-top
quark with angle
$\theta_{\bar{\imath}}$ to the spin axis of the
anti-top quark in the
anti-top rest frame, the correlation is given by
\beq
{1 \over \sigma} ~
{ d^2 \sigma \over d(\cos\theta_i) \thinspace
d(\cos \theta_{\bar{\imath}}) }
{}~ = ~
{1+ \kappa \cos\theta_i \cos \theta_{\bar{\imath}}  \over 4},
\label{doublediff}
\eeq
where
\beq
\kappa = (1-2P_X)\thinspace \alpha_i \alpha_{\bar{\imath}}
\eeq
and $P_X$ is the fractional purity of the unlike-spin component
of the sample of \ttbar\ events.
If both the top and anti-top quark decayed spherically in their
respective rest frames~\cite{spheric},
then the right hand side of eqn (\ref{doublediff}) would be simply
$1\over4$.
Therefore, the contribution
${1\over4}\kappa \cos\theta_i \cos \theta_{\bar{\imath}}$ is
induced by the spin correlations of the \ttbar\ pair.

The strategy to observe these angular correlations
in top production at hadron colliders is to select a sample of
\ttbar\ pairs which have a high asymmetry
in the number of like-spin to unlike-spin pairs,
{\it i.e.}\ dominated by
unlike-spin pairs for the Tevatron or like-helicity pairs for the LHC.
Then, we  choose those events for which the top quark has a given spin
and look to see what the correlations of the decay products
are for the anti-top quark or vice versa.
At the Tevatron if the top quark had spin up, for example,
then the anti-top quark should have spin down,
while at the LHC, if the top quark has right helicity,
then the anti-top quark should also have right helicity.

Suppose we choose those events for which the {\it i\/}th decay
product on the top quark side of the event has an angle
$\theta_i$ in the top rest frame with respect to the
axis defining the top quark spin such that
$\cos \theta_i > y$.   Then, this top quark decay has a
probability $P_{\uparrow}$, given by Eq.~(\ref{Rcut}),
of coming from a spin up top.
Furthermore, on the anti-top quark side of the event,
the $\alpha$ determining the angular correlation of the
{$\bar{\imath}\kern1.0pt$}th
decay product in Eq.~(\ref{alphadefn})
is given by
\beq
	(1-2P_{X})(2P_{\uparrow}-1) \alpha_{\bar{\imath}}.
\label{slope}
\eeq
If we can only determine the identity of the
{$\bar{\imath}\kern1.0pt$}th
decay product probabilistically,
as in the case of the $d$-type quark in hadronic decays,
then $\alpha_{\bar{\imath}}$ in the above expression is replaced by
\beq
P_d \alpha_{d} ~+~ (1-P_d)\alpha_{\bar u},
\eeq
where $P_d$ is given by Eq.~(\ref{Dcut}).

To demonstrate these correlations we choose the $\cos \theta_i$ cut,
which is used to distinguish spin up from spin down,
to be at zero.
This divides the data sample into two sets which we call
spin ``up'' and spin ``down''.
On the other side of the event we can compare the
angular distributions between these two data sets.
Since the charged lepton has the largest correlation
to the spin direction of the quark or anti-quark,
it is natural to use this particle to distinguish spin up from down.
Requiring $\cos\theta > 0$ for the charged lepton yields a
probability of 75\% that it came from a spin up quark,
{\it i.e.} $P_{\up} = 0.75$.
If we tighten this cut to
$\cos\theta > 0.5$, then $P_{\uparrow} =  0.875$,
thus increasing the correlations by 50\%, with a factor
of two loss in statistics.
For dilepton events, the correlations
on the other side of the event which we can study are between
the charged lepton or the $b$-quark and the spin axis, assuming that
the neutrino momenta can be determined.
In the charged lepton plus four jet channel we can look at the
correlations between the {\it ``d''}-type quark or the $b$-quark
and the spin axis.
Here the {\it ``d''}-type quark
is defined as that jet which is closest to the
$b$-quark direction in the $W$-boson rest frame.
This allows us to include all events and is effectively a
$\cos \thetad > 0$ cut.
The probability that this jet comes from a (real) $d$-type quark is
given by eqn~(\ref{Dcut}), and equals 61\% for 175 GeV top quarks.
One further possibility is to look at the correlation between
the $b$-quark and the $\bar{b}$-quark
for all the double-tagged \ttbar\ events,
which may be done in a similar manner.

We have performed a first-pass monte carlo study of these effects
at the parton level without any
hadronization or jet energy smearing effects included.
However, we expect these effects to be small.
Also, we have used the known neutrino
momenta to determine the momenta of the top quarks
and hence the appropriate angles in the top quark rest frames.
Studies by CDF~\cite{gpyeh} demonstrate that even
in dilepton events,
because of the mass constraints on the top quarks and $W$-bosons,
the neutrino momenta can be determined to better than 10\%.
A further complication is the combinatoric background associated
with assigning particles to the wrong top quark decay.
All of these effects would need to be included in a full study of this
phenomena, and would result in a reduction of the correlations
determined below.

\begin{figure}[h]

\vspace*{15cm}
\includegraphics{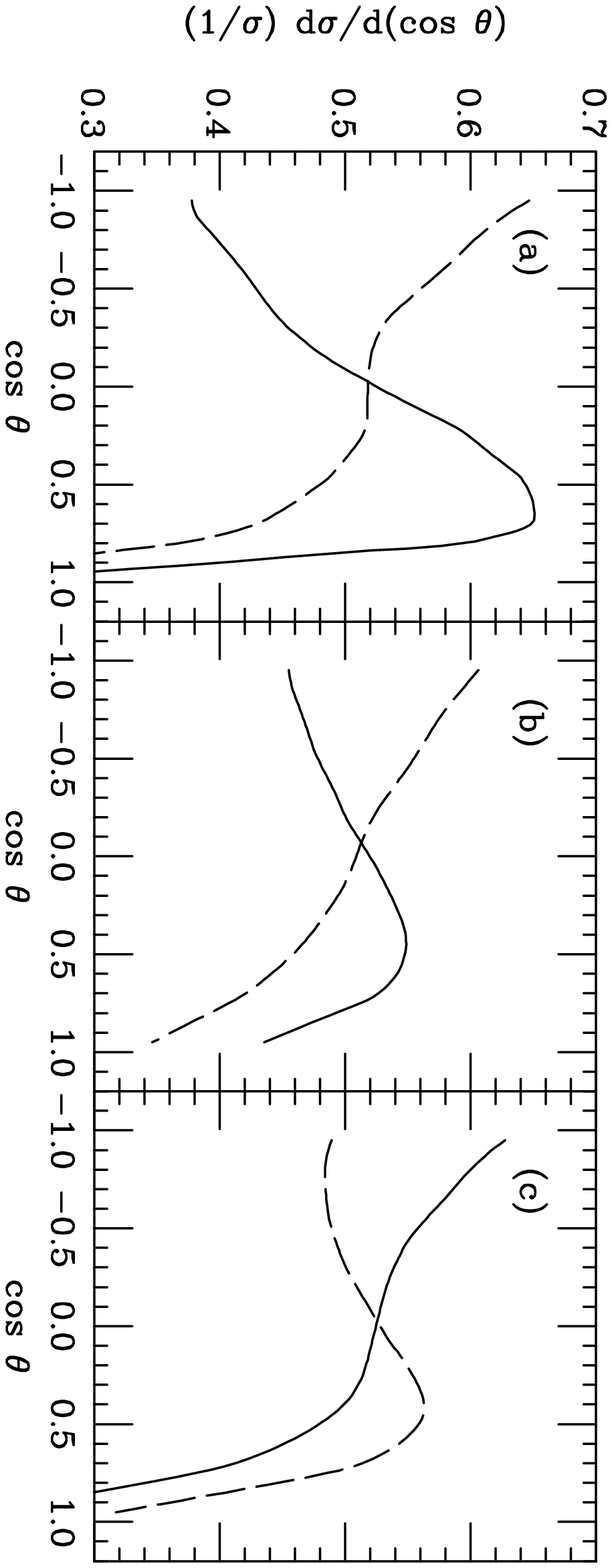}
\vspace{1.0cm}

\caption[]{
Angular correlation between the two charged leptons in
\ttbar\ production and decay.
Plotted is the angle between the charged lepton on the anti-top side
of the event and the $\bar t$ spin axis in the $\bar t$ rest frame.
The data are divided into spin-``up'' (solid) and
spin-``down'' (dashed) top quark components, as determined from the
charged lepton on the top side of the event for
(a) the Tevatron using the \beamline\ basis,
(b) the Tevatron using the helicity basis, and
(c) the LHC using the helicity basis.
}\label{Xmu}
\end{figure}

\begin{figure}[h]

\vspace*{15cm}
\includegraphics{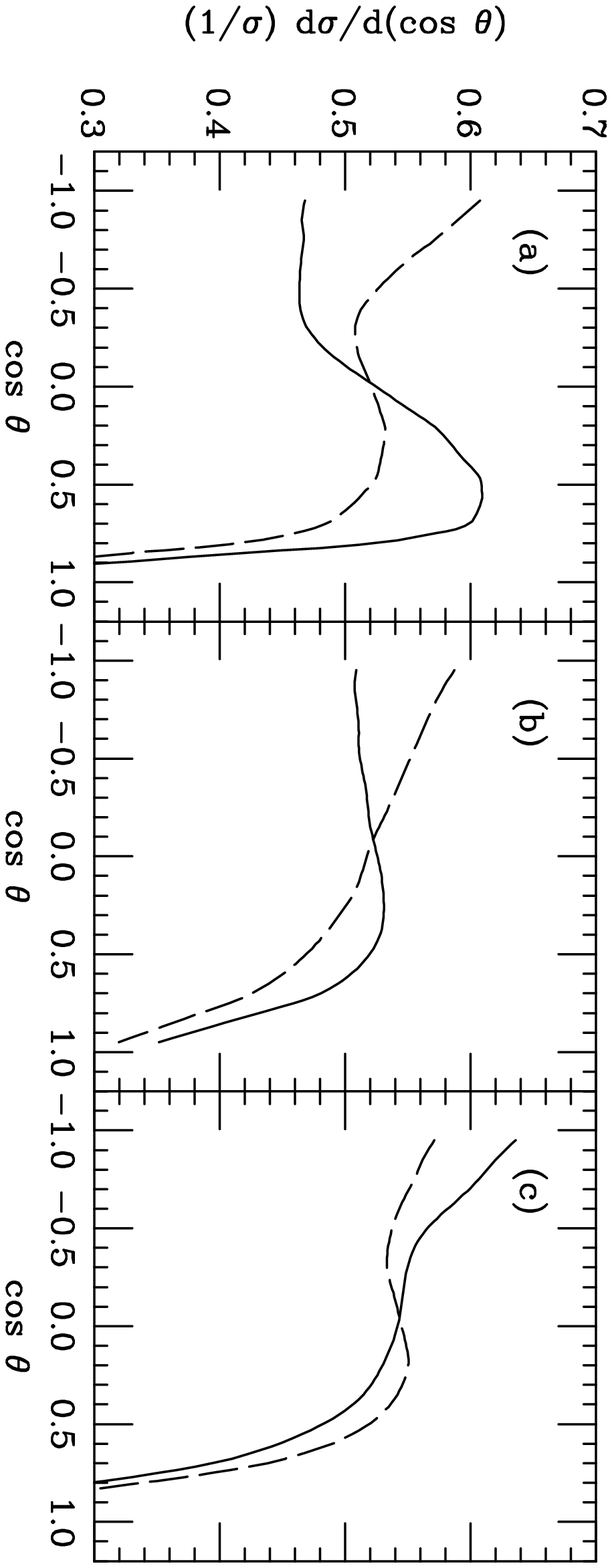}
\vspace{1.0cm}

\caption[]{
Angular correlation between the charged lepton and the
{\it ``d''}-type quark in \ttbar\ production and decay.
Plotted is the angle between the {\it ``d''}-type quark on the
anti-top side of the event and the $\bar t$ spin axis in the
$\bar t$ rest frame.
The data are divided into spin-``up'' (solid) and
spin-``down'' (dashed) top quark components, as determined from the
charged lepton on the top side of the event for
(a) the Tevatron using the \beamline\ basis,
(b) the Tevatron using the helicity basis, and
(c) the LHC using the helicity basis.
}\label{Xd}
\end{figure}

\begin{figure}[h]

\vspace*{15cm}
\includegraphics{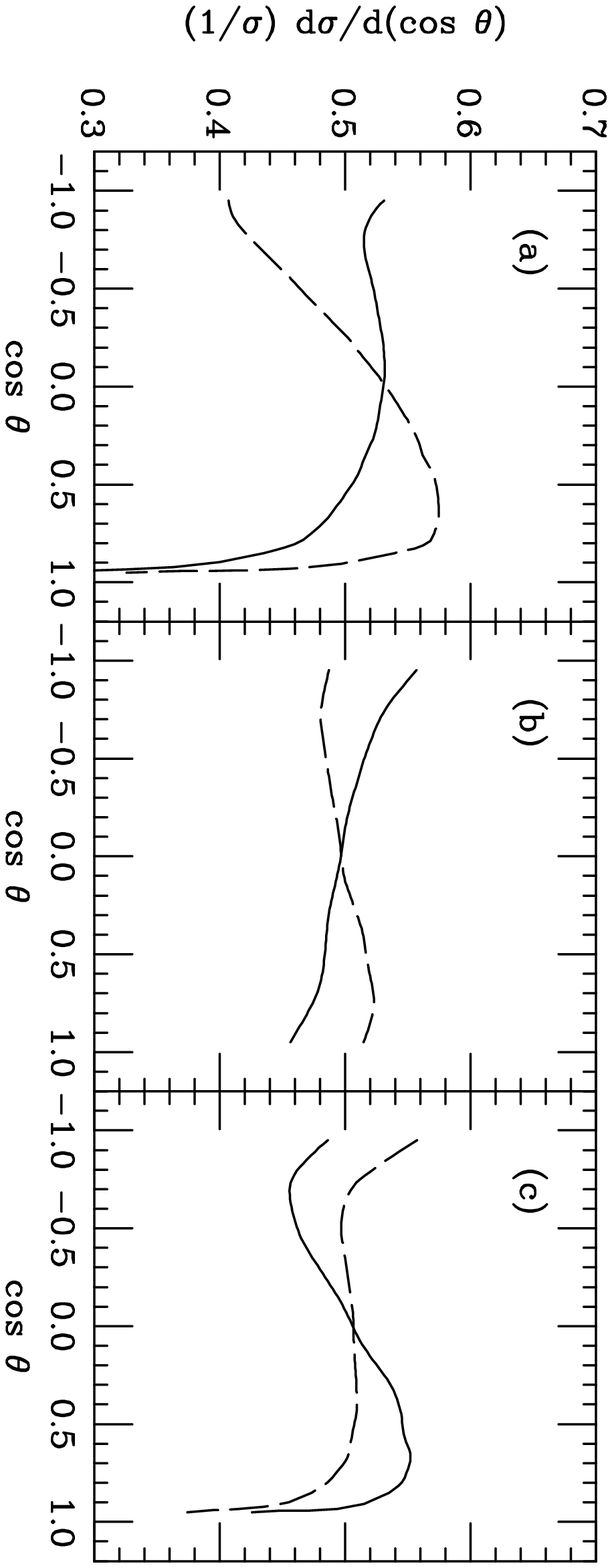}
\vspace{1.0cm}

\caption[]{
Angular correlation between the charged lepton and the
$\bar b$-quark in \ttbar\ production and decay.
Plotted is the angle between the $\bar b$-quark
and the $\bar t$ spin axis in the $\bar t$ rest frame.
The data are divided into spin-``up'' (solid) and
spin-``down'' (dashed) top quark components, as determined from the
charged lepton on the top side of the event for
(a) the Tevatron using the \beamline\ basis,
(b) the Tevatron using the helicity basis, and
(c) the LHC using the helicity basis.
}\label{Xbbar}
\end{figure}

\begin{figure}[h]

\vspace*{15cm}
\includegraphics{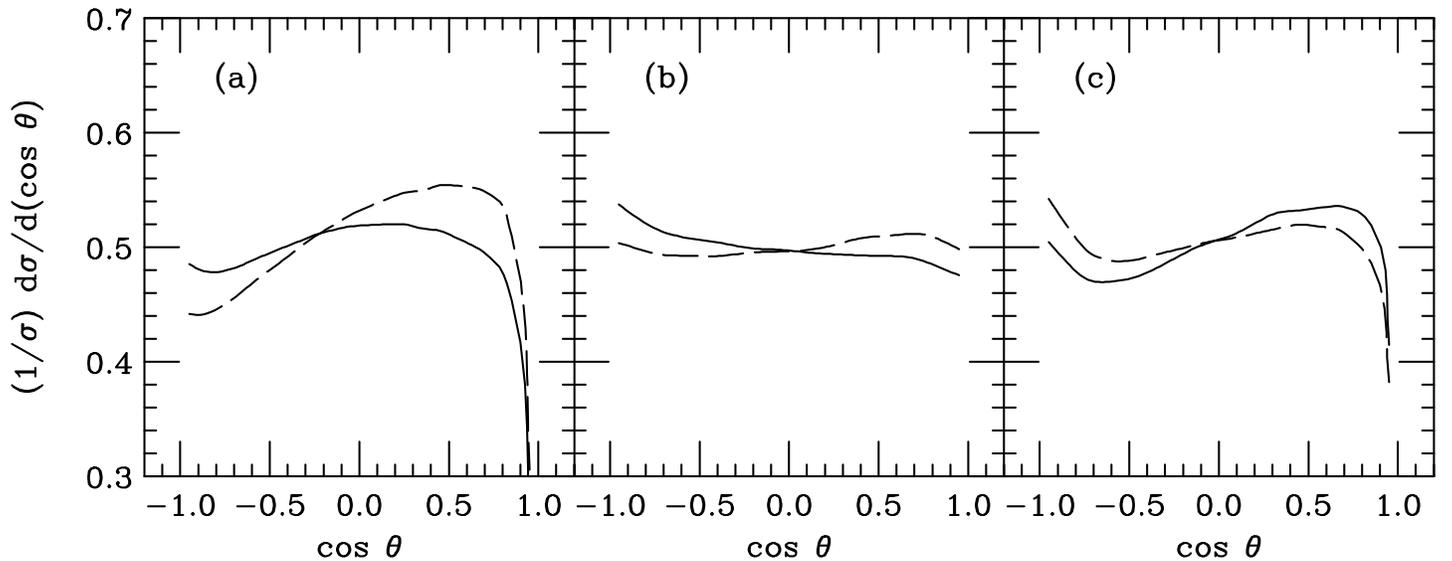}
\vspace{1.0cm}

\caption[]{
Angular correlation between the two $b$-quarks in
\ttbar\ production and decay.
Plotted is the angle between the $\bar b$-quark
and the $\bar t$ spin axis in the $\bar t$ rest frame.
The data are divided into spin-``up'' (solid) and
spin-``down'' (dashed) top quark components, as determined from the
$b$-quark on the top side of the event for
(a) the Tevatron using the \beamline\ basis,
(b) the Tevatron using the helicity basis, and
(c) the LHC using the helicity basis.
}\label{Xb-bbar}
\end{figure}

We selected a \ttbar\ sample for both the Tevatron and the LHC using
the following minimal cuts on the transverse momenta, $p_T$,
and pseudo-rapidities, $\eta$, of all final state particles:
for the Tevatron we required
\beqa
	p_T  >  15{\rm\enspace GeV} , & & \quad
	| \eta |  <  2,
\eeqa
while for the LHC we imposed
\beqa
	p_T  >  20{\rm\enspace GeV}, & & \quad
	| \eta |  <  3.
\eeqa
No further cuts in $M_{t \bar{t}}$ or $\thetas$ were made
to increase the spin asymmetry.
The monte carlo generated events with the full spin correlations
using the Kleiss and Stirling~\cite{ks} matrix elements squared.
The events from this \ttbar\ sample were analyzed
as discussed in the previous paragraph, using the $\cos \theta_i > 0$
selection criteria to divide the sample into two data sets.
In
Figs.~\ref{Xmu}--\ref{Xb-bbar} we compare the results
for four different correlations
in each of the following three cases:
the ``beamline'' basis
at the Tevatron, the helicity basis at the Tevatron, and the
helicity basis at the LHC.
The correlations studied were the charged lepton of one of top quarks
versus the charged lepton, Fig.~\ref{Xmu},
the {\it ``d''}-type quark, Fig.~\ref{Xd},
or the $b$-quark in the other top quark decay, Fig.~\ref{Xbbar},
as well as the correlations between the two $b$-quarks
in the \ttbar\ sample, Fig.~\ref{Xb-bbar}.
For each of the particles at each of the machines we show
the angular distributions of both the spin-``up'' and spin-``down''
data sets using the full spin-correlated
matrix element squared with the minimal cuts.

These plots should be viewed in the light of the following two
observations.
First, if we produce data sets using the
minimal cuts but allow both top quarks to decay spherically in their
respective rest frames~\cite{spheric},
we find that the resulting two curves are identical and
equal to the average of the two curves shown.
Hence, the difference between the plotted curves
comes from the spin correlations
induced in the production of the \ttbar\ pair.
Second, in the absence of the minimal cuts,
the curves
in Figs.~\ref{Xmu}--\ref{Xb-bbar}
would be straight lines going through
$(\thinspace 0.0\thinspace ,\thinspace 0.5\thinspace )$,
with slopes easily calculable from
Eq.~(\ref{slope}).
For the \beamline\ basis at the Tevatron, the $p_T$ and
$\eta$ cuts
are approximately equally important in distorting the
shape of these curves, with the most pronounced effects
around $\cos \theta =1 $.
If we relax
these cuts to $p_T > 10$ GeV and $|\eta| < 3$,
then these curves become nearly equal to the straight lines
of the no cut case, except very close to $\cos \theta = 1$.
For the helicity basis at the Tevatron and the LHC,
the $\eta$ cut plays only a minor role:
for values greater than or equal to 2,
this cut has essentially no effect.
It is the $p_T$ cut which is mainly
responsible for the distortion of these curves from
the ideal straight lines.
The reason that the LHC shows a larger distortion than the Tevatron
is the we have used a higher $p_T$ cut.
Although the center of mass energy at the LHC is seven times
higher than at the Tevatron, top quarks produced at the
LHC have on average only 10--20\%  higher $p_T$.
Therefore, the same $p_T$ cut has nearly the same effect
on the correlations at both machines.  For $p_T$ cuts very
much above 20 GeV, the distortions become unacceptably large,
and the two curves are forced closer and closer together.

After sufficient \ttbar\ events have been collected by the
Tevatron or the LHC, the difference between the ``up'' and ``down''
data sets could be enhanced
by making extra cuts to increase the spin asymmetry and/or
tightening the
selection criteria on what we have have called spin
``up'' and ``down'' top quarks.

%%%%%%%%%%%%%%%%%%%%%%%%%%%%%%%%%%%%%%%%%%%%%%%%%%%%%%%%%%%%%%%%
%%
%%      NEW PHYSICS EFFECTS
%%
%%%%%%%%%%%%%%%%%%%%%%%%%%%%%%%%%%%%%%%%%%%%%%%%%%%%%%%%%%%%%%%%

\section{Signatures of New Physics} \label{NEWPHYS}

In this section we briefly discuss the effects of
new physics on the correlations examined in the previous section.
Hill and Parke \cite{hp} have proposed that the production of top pairs
at hadron colliders could be affected by a new vector
particle associated
with top-color. Such a resonance would appear in
the angular correlations
for top pair production by changing the relative mixture
of $q\bar q$ to
$gg$ initiated production of top quarks, and by distorting the zero
momentum frame speed, $\beta$, for the $q\bar q$ component.
At the Tevatron both of these effects would increase the
LR+RL helicity component in top pair production so as to
increase the correlations discussed in the previous section.
Since the $q\bar q$ component at the LHC is a small fraction of the
total cross section, small changes in this component
will be difficult to see.

Eichten and Lane \cite{el} have discussed the
effects of a techni-eta in two
scale technicolor on top quark pair production at hadron colliders.
Since the production of top pairs via a scalar or pseudoscalar goes
exclusively into
the LL+RR helicity state, the effect at the
Tevatron of such a resonance is to reduce the
correlations discussed in this paper. If the techni-eta has a mass just
above the top pair threshold the charged
leptons in the dilepton events will
tend to be in the same hemisphere instead of in opposite hemispheres.
At the LHC such a resonance would enhance
the correlations produced by the
Standard Model $gg$ component.

On the decay end, new physics such as a charged Higgs
decay of top would also affect these correlations.
The correlation coefficients $\alpha$ for the
decay $t \rightarrow b jj$ via a charged Higgs have values of
$\alpha_b = 1.0$ and  $\alpha_j=( -\xi^2+1+2\xi \ln \xi)/(\xi-1)^2$,
where $\xi=m_t^2/m_H^2$.
As a result, a deviation in the standard model correlations
in the $W$ plus four jet sample of top pair production
could be observed if the branching fraction for top
into charged Higgs plus $b$-quark is large enough.

%%%%%%%%%%%%%%%%%%%%%%%%%%%%%%%%%%%%%%%%%%%%%%%%%%%%%%%%%%%%%%%%
%%
%%      CONCLUSIONS
%%
%%%%%%%%%%%%%%%%%%%%%%%%%%%%%%%%%%%%%%%%%%%%%%%%%%%%%%%%%%%%%%%%

\section{Conclusions} \label{CONCLUSIONS}

We have described a method whereby the angular
correlations between the
top and anti-top decay products could be observed at a hadron
collider.
Our discussion is based upon the asymmetry in the
number of like-spin to unlike-spin
\ttbar\ pairs produced at any hadron collider.
When the production
is dominated by quark-antiquark annihilation, there will be
an excess in the number of unlike-spin \ttbar\ pairs using the
\beamline\ basis and unlike-helicity pairs using the helicity basis.
On the other hand, when gluon-gluon fusion dominates the production,
there will be an excess of like-helicity pairs.
The size of
these excesses may be enhanced by applying a cut on any variable
that selects events in a restricted $\beta$ region in the zero
momentum frame of the \ttbar\ pair.
The spin of a given
top quark may be determined probabilistically by
considering the angle between
the direction of motion of the decay products
and the direction of the spin axis.
The charged lepton or $d$-type quark from the $W$-boson decay
have the highest correlations to the top quark spin axis.
If we use these correlations to divide the data into a spin ``up'' and
spin ``down'' component for the top quark, we can observe a difference
between these two data sets in the angular correlations between the
anti-top spin axis
and the direction of motion of the anti-top decay products.
For a ``loose'' set of cuts,
we find that the difference between the
correlations for the spin ``up''
verses spin ``down'' data samples can be as large as 25\%
at the Tevatron and
14\% at the LHC,
making these effects potentially observable.
If the top quark is strongly coupled
to new physics beyond the Standard Model,
then these correlations could be dramatically altered.

%%%%%%%%%%%%%%%%%%%%%%%%%%%%%%%%%%%%%%%%%%%%%%%%%%%%%%%%%%%%%%%%
%%
%%      ACKNOWLEDGEMENTS
%%
%%%%%%%%%%%%%%%%%%%%%%%%%%%%%%%%%%%%%%%%%%%%%%%%%%%%%%%%%%%%%%%%

\acknowledgements

The Fermi National Accelerator
Laboratory is operated by Universities Research Association,
Inc., under contract DE-AC02-76CHO3000 with the U.S. Department
of Energy.
High energy physics research at the University of Michigan
is supported in part by the U.S. Department of Energy,
under contract DE-FG02-95ER40899.
SP would like to thank T.~Liss, R.~Raja, A.~Tollestrup
and G.~P.~Yeh for valuable discussions.
GDM would like to thank the Aspen Center for
Physics, where a portion of this work was completed.
% The computer on which the monte carlos were generated
% runs the Linux operating system. Thanks to Linus Torvalds
% and his supporting cast of thousands for creating it.

%%%%%%%%%%%%%%%%%%%%%%%%%%%%%%%%%%%%%%%%%%%%%%%%%%%%%%%%%%%%%%%%
%%
%%      APPENDIX
%%
%%%%%%%%%%%%%%%%%%%%%%%%%%%%%%%%%%%%%%%%%%%%%%%%%%%%%%%%%%%%%%%%

\newpage
\appendix

\section*{Spinor Helicity Basis for Massive Fermions}

In this appendix, we discuss the spinor helicity basis for
massive fermions used to derive many of the results contained
in this paper.
This appendix follows the conventions and notation used in the
review by Mangano and Parke \cite{mp},
and is a very useful extension to that review.
The connection to the spin state methods found in
Bjorken and Drell \cite{BjD} (\BD) is also included.

For a massive particle of momentum $P$ and mass $M$,
we follow Kleiss and Stirling~\cite{ksspinors} and
pick a reference vector, $p_2$,
which is lightlike, $p_2^2=0$.
Usually it is convenient to choose $p_2$ as one of the
massless particles in the
situation under consideration.
As we shall see later, the direction opposite to the spatial
momentum of $p_2$
in the rest frame of the massive particle defines the axis along
which the spin of the massive particle is decomposed.
Then we define the vector $p_1$ by
$ p_1 =  P - \frac{M^2}{2 P \cdot p_2} ~p_2 $.
Note that $p_1$ is also a massless vector, $p_1^2=0$,
and that
\beq
P=p_1 + \frac{M^2}{2 p_1 \cdot p_2} p_2.
\eeq
For some applications it is convenient to rescale $p_2$
so that $P=p_1+p_2^{\prime}$,
where $p_2^{\prime}= \frac{M^2}{2 p_1 \cdot p_2}~p_2$.

To obtain the spinors which are eigenstates
of spin for the massive particle,
we need to define two complex square roots of the factor
$ \frac{M^2}{2 p_1 \cdot p_2} $ by
\beq
	\alpha  \equiv  \frac{M}{\sppr{p_1-}{p_2+}},
\quad \quad	\beta   \equiv  \frac{M}{\sppr{p_2+}{p_1-}}.
\eeq
With these definitions
$\alpha \beta  =  \frac{M^2}{2 p_1 \cdot p_2}$.

Then, the basis spinors describing the massive particle spin states are
\beqa
	&	u_{\uparrow}(P) = \rsp{p_1+} - \beta ~\rsp{p_2-}
\quad\quad &    u_{\downarrow}(P)  = \rsp{p_1-} + \alpha ~\rsp{p_2+}
\label{u} \\
	&	v_{\uparrow}(P)  = \rsp{p_1-} - \alpha ~\rsp{p_2+}
\quad\quad &	v_{\downarrow}(P)  = \rsp{p_1+} + \beta ~\rsp{p_2-}
\label{v} \\
	&      \bar{u}_{\uparrow}(P)  = \lsp{p_1+} - \alpha ~\lsp{p_2-}
\quad\quad &   \bar{u}_{\downarrow}(P)  = \lsp{p_1-} + \beta ~\lsp{p_2+}
\label{ubar} \\
	&      \bar{v}_{\uparrow}(P) = \lsp{p_1-} - \beta ~\lsp{p_2+}
\quad\quad &  \bar{v}_{\downarrow}(P) = \lsp{p_1+} + \alpha ~\lsp{p_2-}.
\label{vbar}
\eeqa
As expected, the spin states are
a superposition of the two possible chiralities.
They satisfy all the usual relations:\\
\indent the Dirac equations
\beqa
&	(\negthinspace\Pslash-M) u(P)  = 0, \quad \quad
& \bar{u}(P)(\negthinspace\Pslash - M) = 0,
\label{DiracEqn} \\
&       (\negthinspace\Pslash+M) v(P)  = 0, \quad \quad
& \bar{v}(P)(\negthinspace\Pslash + M) = 0,
\eeqa
\indent
the completeness conditions
\beqa
&	\displaystyle\sum_{\lambda} u_{\lambda}(P) \bar{u}_{\lambda}(P)
       = \Pslash + M,
\quad   \quad
&	\sum_{\lambda} v_{\lambda}(P) \bar{v}_{\lambda}(P)
       =\Pslash - M,
\eeqa
\indent
and the orthogonality conditions
\beqa
	\bar{u}_{\lambda_1}(P) u_{\lambda_2}(P)=&
                            2M\delta_{\lambda_1,\lambda_2},
\quad\quad
	\bar{v}_{\lambda_1}(P) v_{\lambda_2}(P)=&
                           -2M\delta_{\lambda_1,\lambda_2} ,
\label{Normalization} \\
	\bar{u}_{\lambda_1}(P) v_{\lambda_2}(P)=&0  ,
\quad\quad\quad\quad\quad\enspace
	\bar{v}_{\lambda_1}(P) u_{\lambda_2}(P)=&  0.
\label{Orthogonal}
\eeqa

To make contact with the methods of \BD\
for massive fermion states, we must first make trivial modifications
to account for \BD's choice of normalization.
Instead of decomposing the
particle's momentum, however, \BD\ make use of a vector
$s$, which satisfies
\beq
s^2 = -1, \quad P\cdot s = 0.
\eeq
In the particle's rest frame, the spatial part of $s$
points in the same direction as the particle's spin.
The relation between the two descriptions is provided by the
following identities
\beq
p_1  =  {{P+Ms}\over{2}},
\quad \quad \frac{M^2}{2 p_1 \cdot p_2} p_2  =  {{P-Ms}\over{2}}.
\label{SpinConversion}
\eeq
To see that this is indeed correct,
evaluate some outer products $u\bar u$ or $v \bar v$ for some spin
projection using our spinors,
make the above substitutions, and you will recover the \BD\ expressions,
{\it e.g.}\
$u(P,s) \bar u(P,s) = {1\over2}({\Pslash} + M)(1 + \gamma_5{\sslash})$.

To describe the spin direction in terms of $p_2$, we
invert~(\ref{SpinConversion}), and evaluate the resulting expression
in the rest frame of the massive particle,
where $p_2$ points in the direction of some unit vector $\nhat$.
For $P = (M, \vec{0})$
and $p_2 = (1,\nhat)$ we obtain $s = (0, -\nhat)$.
Therefore, the direction of
the particle's spin is {\it opposite} to the direction of the spatial
part of $p_2$
\cite{energy}.
Alternatively, the particle's spin is in the {\it same} direction as
the spatial part of $p_1$ in the massive particle's rest frame.

Next, we consider eigenstates of helicity.
Since helicity is
simply the spin projected along
the direction of
motion of the particle, choose
$p_2 = p(1, -\nhat)$
for a massive particle with momentum
$P = ( \sqrt{p^2+M^2}, p ~\nhat )$.
Then,
\beqa
p_1 & = & \frac{\sqrt{p^2+M^2}+p}{2}~(1,\nhat)  \\[0.2in]
\frac{M^2}{2 p_1 \cdot p_2} ~p_2 & = &
\frac{\sqrt{p^2+M^2}-p}{2}~(1,-\nhat) .
\eeqa
It is conventional to label these helicity states
by ``L'' and ``R'', instead
of ``$\downarrow$'' and ``$\uparrow$'' respectively.

In the large momentum limit,
\beq
p_1  \rightarrow  P, \quad \quad
\frac{M^2}{2 p_1 \cdot p_2} ~p_2  \rightarrow   0.
\label{Plimit}
\eeq
Therefore the basis spinors
(\ref{u})--(\ref{vbar}) become
pure chirality eigenstates:
\beqa
	&	u_{\UP}(P) \rightarrow \rsp{p_1+}
\quad	&	u_{\DOWN}(P)  \rightarrow \rsp{p_1-}
\label{ulimit} \\
	&	v_{\UP}(P)  \rightarrow \rsp{p_1-}
\quad	&	v_{\DOWN}(P)  \rightarrow \rsp{p_1+}
\label{vlimit} \\
	&	\bar{u}_{\UP}(P)  \rightarrow \lsp{p_1+}
\quad	&	\bar{u}_{\DOWN}(P)  \rightarrow \lsp{p_1-}
\label{ubarlimit} \\
	&	\bar{v}_{\UP}(P) \rightarrow \lsp{p_1-}
\quad	&	\bar{v}_{\DOWN}(P) \rightarrow \lsp{p_1+} .
\label{vbarlimit}
\eeqa
Thus, in what is equivalent to the massless limit,
the right-handed helicity state $u_{\UP}$ becomes
a state of pure right-handed
chirality.  That the right-handed helicity state $v_{\UP}$
becomes a state of pure {\it left}-handed chirality simply
reflects the fact that the helicity and chirality eigenvalues
are {\it opposite}\ in sign for the anti-particle.

%%%%%%%%%%%%%%%%%%%%%%%%%%%%%%%%%%%%%%%%%%%%%%%%%%%%%%%%%%%%%%%%
%%
%%      REFERENCES
%%
%%%%%%%%%%%%%%%%%%%%%%%%%%%%%%%%%%%%%%%%%%%%%%%%%%%%%%%%%%%%%%%%

%%%%%%%%%%%%%%%%%%%%%%%%%%%%%%%%%%%%%%%%%%%%%%%%%%%%%%%%%%%%%%%%
%%
%%      FIGURE CAPTIONS
%%
%%%%%%%%%%%%%%%%%%%%%%%%%%%%%%%%%%%%%%%%%%%%%%%%%%%%%%%%%%%%%%%%
\vspace*{1cm}

%%%%%%%%%%%%%%%%%%%%%%%%%%%%%%%%%%%%%%%%%%%%%%%%%%%%%%%%%%%%%%%%
%%
%%      TABLES
%%
%%%%%%%%%%%%%%%%%%%%%%%%%%%%%%%%%%%%%%%%%%%%%%%%%%%%%%%%%%%%%%%%

\begin{table}
\caption{Correlation coefficients $\alpha$ for both semi-leptonic
and hadronic top quark decays as a function of
$\xi\equiv m_t^2/m_w^2$ in the narrow width approximation
for the $W$-boson and using $m_b=0$. For $m_t > 100$ GeV these
are excellent approximations.
\label{Alphas}}
\begin{tabular}{cccccccc}
&&&$\bar e$ or $\bar d$&  $1$&&& \\[0.1in]
&&&$\nu$ or $u$ &  $\displaystyle{  { (\xi-1)(\xi^2 - 11\xi - 2)
                               + 12\xi \ln \xi}
                             \over
                             { (\xi+2)(\xi-1)^2 } }$&&& \\[0.1in]
&&&$W^+$ &  $\displaystyle{ {\xi-2}\over{\xi+2} }$&&& 	\\[0.1in]
&&&$b$ &  $\displaystyle{ - ~{{\xi-2}\over{\xi+2}} }$&&&
\end{tabular}
\end{table}


\begin{references}
\bibitem{NOslashes}  To make these results easier to read the slashes
have been left off the momentum vectors in the traces.
\bibitem{spheric}
By spherical decay we mean that the decay
$t \rightarrow b~W^+$ is spherical in the $t$
rest frame. However, the subsequent
decay of the $W$-boson has all of the Standard Model
spin correlations included.
\bibitem{energy}
None of above definitions depend on the sign of the energy component of
any of the vectors. However if the massive particle and $p_2$ have
opposite sign of their energy components then the spin axis is in
the same direction as the spatial momentum of $p_2$.
\end{references}
\end{document}